\documentclass[12pt]{article}
\pdfoutput=1
\usepackage{putex}
\usepackage{graphicx}
\usepackage{epstopdf}
\usepackage{enumerate}
\usepackage{cite}
\usepackage{tensor}
\usepackage{slashed}

\numberwithin{equation}{section}

\newcommand{\abs}[1]{\left\lvert #1 \right\rvert}

\newcommand {\be} {\begin {equation}}
\newcommand {\ee} {\end {equation}}

\newcommand {\bes} {\begin {equation*}}
\newcommand {\ees} {\end {equation*}}

\newcommand{\es}[2] {\begin{equation} \label{#1} \begin{split} #2 \end{split} \end{equation}}

\newcommand{\Z}{\mathbb{Z}}
\newcommand{\N}{\mathbb{N}}
\newcommand{\R}{\mathbb{R}}

\def\Tr{\mop{Tr}}

\newcommand{\beq}{\begin{equation}}
\newcommand{\eeq}{\end{equation}}

\begin{document}

\institution{IAS}{School of Natural Sciences, Institute for Advanced Study, Princeton, NJ 08540}
\institution{MIT}{Center for Theoretical Physics, Massachusetts Institute of Technology, Cambridge, MA 02139}
\institution{Harvard}{Department of Physics, Harvard University, Cambridge, MA 02138}
\institution{PU}{Joseph Henry Laboratories, Princeton University, Princeton, NJ 08544}

\title{Entanglement Entropy of 3-d Conformal Gauge Theories with Many Flavors}

\preprint{PUPT-2402\\MIT-CTP-4337}

\authors{Igor R.~Klebanov,\worksat{\IAS}\footnote{On leave from
Joseph Henry Laboratories and Center for Theoretical Science, Princeton University.}
  Silviu S.~Pufu,\worksat{\MIT} Subir Sachdev,\worksat{\Harvard}
Benjamin R.~Safdi\worksat{\PU}}

\abstract{
Three-dimensional conformal field theories (CFTs) of deconfined gauge fields coupled to gapless flavors of fermionic and bosonic matter describe quantum critical points of condensed matter systems in two spatial dimensions.  An important characteristic of these CFTs is the finite part of the entanglement entropy across a circle.  The negative of this quantity is equal to the finite part of the free energy of the Euclidean CFT on the three-sphere, and it has been proposed to satisfy the so called $F$-theorem, which states that it decreases under RG flow and is stationary at RG fixed points.  We calculate the three-sphere free energy of non-supersymmetric gauge theory with a large number $N_F$ of bosonic and/or fermionic flavors to the first subleading order in $1/N_F$.  We also calculate the exact free energies of the analogous chiral and non-chiral ${\cal N} = 2$ supersymmetric theories using localization, and find agreement with the $1/N_F$ expansion. We analyze some RG flows of supersymmetric theories, providing further evidence for the $F$-theorem.
}

\date{December 2011}

\maketitle


\section{Introduction}
\label{sec:intro}

Many interesting quantum critical points of condensed matter systems in two spatial dimensions \cite{Wen:1993zza, Chen:1993cd, ssqhe, rw1, rw2, Motrunich:2004zz, senthil1, senthil2, hermele1, hermele2, ran, acl, Kaul:2008xw, sstasi} are described by conformal field theories (CFT) in three spacetime dimensions where massless fermionic and/or bosonic matter interacts with deconfined gauge fields. These include critical points found in insulating antiferromagnets and $d$-wave superconductors and between quantum Hall states.  Such CFTs can be naturally analyzed by an expansion in $1/N_F$, where $N_F$ is the number of `flavors' of matter. This large $N_F$ limit is taken at fixed $N_c$, where $N_c$ is a measure of the size of the gauge group {\em e.g.} the non-abelian gauge group $U(N_c)$. Classic examples of such CFTs include three-dimensional $U(1)$ gauge theory coupled to a large number of massless charged scalars \cite{Appelquist:1981vg} or Dirac fermions \cite{Appelquist:1986fd,Appelquist:1988sr}. These theories are conformal to all orders in the $1/N_F$ expansion, and they are widely believed to be conformal for $N_F > N_\text{crit}$, where $N_\text{crit}$ is a conjectured critical number of flavors dependent on the choice of the gauge group \cite{Appelquist:1986fd,Appelquist:1988sr}.  The 3-dimensional CFTs may also contain Chern-Simons terms whose coefficients $k$ may be taken to be large.

An important characteristic of a 3-dimensional CFT is the ground state entanglement entropy across a circle of radius $R$. Its general structure is
\beq
S = \alpha { R \over \epsilon}  - F \,,
\label{genent}
\eeq
where $\epsilon$ is the short distance cut-off. As established in \cite{ch2} (see also \cite{Casini:2010kt,ch1}) the subleading $R$-independent term is related to the regulated Euclidean path integral $Z$ of the CFT
on the three-dimensional sphere $S^3$: $F=-\log |Z|$. The quantity $F$ has been conjectured to decrease along any RG flow
\cite{sinha,Jafferis:2011zi,ch2,Klebanov:2011gs}.\footnote{After the original version of this paper appeared, a proof of the F-theorem was presented in \cite{Casini:2012ei}.} This conjecture was inspired by the $c$-theorem in two spacetime dimensions
\cite{Zamolodchikov:1986gt} and the $a$-theorem in four spacetime dimensions \cite{Cardy:1988cwa,Komargodski:2011vj,Komargodski:2011xv}.

In any 3-dimensional field theory with ${\cal N}\geq 2$ supersymmetry, the $S^3$ free energy $F$ may be calculated using the method of localization \cite{Kapustin:2009kz,Drukker:2010nc,Jafferis:2010un,Festuccia:2011ws}.
It has also been calculated in some simple non-supersymmetric CFTs, such as free field theories \cite{Casini:2010kt,ch1,Dowker:2010yj,Klebanov:2011gs} and the Wilson-Fisher fixed point of the $O(N)$ model for large $N$ \cite{Klebanov:2011gs}, which has been conjectured \cite{Klebanov:2002ja} to be dual to Vasiliev's
higher-spin gauge theory in $AdS_4$ \cite{Vasiliev:1995dn}.
In this paper we present the calculation of $F$ in certain 3-d gauge theories coupled to a large number of massless flavors, to the first subleading order in $1/N_F$. We will find that this subleading term is of order $\log N_F$.

The CFTs we study have the following general structure. The matter sector has Dirac fermions $\psi_\alpha$, $\alpha = 1 \dots N_f$,
and complex scalars, $z_a$, $a = 1 \ldots N_b$. We will always take the large $N_f$ limit with $N_b/N_f$ fixed, and use the symbol $N_F$
to refer generically to either $N_f$ or $N_b$. These matter fields
are coupled to each other and a gauge field $A_\mu$ by a Lagrangian of the form
 \es{Lm}{
\mathcal{L}_m = \sum_{\alpha=1}^{N_f} \overline{\psi}_\alpha \gamma^\mu D_\mu \psi_\alpha + \sum_{a=1}^{N_b}
\bigg ( |D_\mu z_a|^2 + s |z_a|^2 + \frac{u}2 \left( |z_a|^2 \right)^2 \bigg ) + \ldots \,,
 }
where $D_\mu = \partial_\mu - i A_\mu$ is the gauge covariant derivative, and the ellipses represent additional possible contact-couplings
between the fermions and bosons, such as a Yukawa coupling. The scalar ``mass'' $s$ generally has to be tuned to reach
the quantum critical point at the renormalization group (RG) fixed point, which is described by a three dimensional CFT; however, this is the only relevant perturbation at the CFT fixed point, and so only a single parameter has to be tuned to access the fixed point. In some cases,
such scalar mass terms are forbidden, and then the CFT describes a quantum critical phase. All other couplings, such as $u$ and the Yukawa coupling, reach values associated with the RG fixed point, and so their values are immaterial for the universal properties of interest in the present paper.

The gauge sector of the CFT has a traditional Maxwell term, along with a possible Chern-Simons term
 \es{LA}{
\mathcal{L}_A =  \frac{1}{2 e^2} \Tr F^2 + \frac{i k}{2 \pi} \Tr \left( F \wedge A  - \frac{1}{3} A \wedge A \wedge A \right) \,.
 }
The gauge coupling $e^2$ has dimension of mass in three spacetime dimensions. It flows to an RG fixed point value, and so its value is also immaterial; indeed, we can safely take the limit $e^2 \rightarrow \infty$ at the outset.
However, our results will depend upon the value of the Chern-Simons coupling $k$, which is RG invariant. We will typically take the large $N_F$ limit with $k/N_F$ fixed at fixed $N_c$, and in most of this paper we set $N_c=1$ for simplicity. (This is to be contrasted with the
`t Hooft type limit of large $N_c$ where $k/N_c$ is held fixed; see, for example, recent work \cite{Minwalla:2011ma,Giombi:2011kc,Aharony:2011jz}.) One of our principal results, established in section~\ref{nonSUSY}, is that for the $U(1)$ gauge theory with Chern-Simons level $k$, coupled to
$N_f$ massless Dirac fermions and $N_b$ massless complex scalars of charge $1$ as in (\ref{Lm}) with $s=u=0$,
\es{prinres}{
F  = \frac{\log 2}{4}\left( N_f  + N_b \right) + \frac{3\zeta(3)}{8 \pi^2} (N_f - N_b)+
\frac{1}{2} \log\left[ \pi \sqrt{ \left(\frac{N_f + N_b}{8}\right)^2 + \left( \frac{k}{\pi} \right)^2} \right] +\ldots \,.
 }
This formula shows that the entanglement entropy is not simply the sum of the topological
contribution $-\frac 12 \log k$ and the contribution of the gapless bulk modes, unlike in the models of
\cite{Swingle:2011ww}.  For CFTs with interacting scalars relevant for condensed matter applications, we have to consider the $u \rightarrow \infty$ limit, and this yields a correction of order unity, with $F \rightarrow F - \zeta(3) / (8 \pi^2)$ at this order \cite{Klebanov:2011gs}. All higher order corrections to \eqref{prinres} are expected
to be suppressed by integer powers of $1/N_F$, whose coefficients do not contain any factors of $\log N_F$.

In section~\ref{SUSYscn} we
will examine similar ${\cal N}= 2$ supersymmetric CFTs on $S^3$ using the localization approach.  We consider
theories with chiral and non-chiral flavorings.
 The partition function $Z$ on $S^3$ is given by a finite-dimensional integral, which has to be locally minimized with respect to the scaling dimensions of the matter fields \cite{Jafferis:2010un}. For a theory with $N$ charged superfields we develop $1/N$ expansions for the scaling dimensions and for the entanglement entropy. As for the non-supersymmetric case, the subleading term in $F$ is of order $\log N$. The coefficient of this term computed via localization agrees with the direct perturbative calculation (\ref{prinres}).

In the supersymmetric case it is possible to develop the $1/N$ expansions to a rather high order, and we compare them with
 precise numerical results. This comparison yields an unprecedented test
of the validity and accuracy of the $1/N$ expansion. At least for supersymmetric CFTs, we find the $1/N$ expansion is
accurate down to rather small values of $N$. We also note a recent numerical study \cite{kaul2}, which found reasonable accuracy
in the $1/N_b$ expansion for a non-supersymmetric CFT.

\section{Mapping to $S^3$ and large $N_F$ expansion} \label{largeNf}

Let us start by examining the case of a $U(1)$ gauge field.  After sending $e^2 \to \infty$, the combined Lagrangian ${\cal L}_m + {\cal L}_A$ obtained from \eqref{Lm} and \eqref{LA} contains two relevant couplings $s$ and $u$, and we should first understand to what values we need to tune them in order to describe an RG fixed point.  Let's ignore for the moment the fermions and the gauge field and focus on the complex scalar fields.  The path integral on a space with arbitrary metric is
 \es{ZScalars}{
  Z = \int Dz_a \exp\left[- \int d^3r  \, \sqrt{g} \left( |\partial_\mu z_a|^2 + s |z_a|^2 + \frac u2 \left( |z_a|^2 \right)^2 \right) \right] \,.
 }
With the help of an extra field $\lambda$, this path integral can be equivalently written as
  \es{ZScalarsEquiv}{
  Z = {\cal C} \int Dz_a D\lambda \exp\left[- \int d^3r  \, \sqrt{g} \left( |\partial_\mu z_a|^2 + s |z_a|^2 - i \lambda |z_a|^2 + \frac 1{2u} \lambda^2 \right) \right] \,,
 }
where the normalization factor ${\cal C}$ defined through ${\cal C} \int D\lambda \exp\left[- \int d^3r  \, \sqrt{g} \left( \frac 1{2u} \lambda^2 \right) \right] = 1$ was introduced so that the value of the path integral stays unchanged.

In flat three-dimensional space, we can tune $s = u = 0$ and describe a non-interacting CFT of $N_b$ complex scalars.  If instead we tune $s = 0$ and send $u \to \infty$, the path integrals \eqref{ZScalars} and \eqref{ZScalarsEquiv} describe the interacting fixed point that we will primarily be interested in in this paper.
We can also send both $s$ and $u$ to infinity, in which case the path integrals above describe the empty field theory.  Using conformal symmetry, we can map each of these fixed points to $S^3$ by simply mapping all the correlators in the theory.  Indeed, since the metric on $S^3$ is equal to that on $\R^3$ up to a conformal transformation,
 \es{S3Metric}{
  ds^2_{S^3} = \frac{4}{\left(1 + \abs{\vec{r}}^2\right)^2} ds_{\R^3}^2 \,,
 }
the mapping of correlators to $S^3$ is achieved by replacing
 \es{Replacement}{
  \vec{r} - \vec{r}' \to \frac{2 (\vec{r} - \vec{r}')}{\left(1 + \abs{\vec{r}}^2 \right)^{1/2} \left(1 + \abs{\vec{r}'}^2 \right)^{1/2}}
 }
in all the flat-space expressions.\footnote{This replacement certainly works for correlators of scalar operators.  In the case of vector operators it is still true that one can use \eqref{Replacement} provided that the $S^3$ correlators are expressed in a frame basis, as in the following section.}  While the theory on $S^3$ defined this way certainly has the correlators of a CFT, it may be a priori not clear which action, and in particular which values of $s$ and $u$, one should choose in order to reproduce these correlators.

In order to study the free theory on $S^3$ one should tune $s = 3/4$ and $u= 0$.  This result holds to all orders in $N_b$ and one can understand it as follows.  The two-point connected correlator of $z_a$ on $\R^3$ is
 \es{TwoPointR3}{
  \langle \bar z_a(r) z_b(r') \rangle_\text{free}^{\R^3} = \frac{\delta_{ab}}{4 \pi \abs{\vec{r} - \vec{r'}}} \,,
 }
because it is the unique solution to the equation of motion following from \eqref{ZScalars} with a delta-function source, $-\nabla^2_{\R^3}  \langle \bar z_a(r) z_b(r') \rangle_\text{free}^{\R^3} = \delta_{ab} \delta^{(3)}(\vec{r} - \vec{r'})$.  Using the mapping \eqref{Replacement} we infer that the corresponding two-point correlator on $S^3$ should be
 \es{TwoPointS3}{
   \langle \bar z_a(r) z_b(r') \rangle_\text{free}^{S^3} = \frac{\delta_{ab} \left(1 + \abs{\vec{r}}^2 \right)^{1/2} \left(1 + \abs{\vec{r}'}^2 \right)^{1/2}}{8 \pi \abs{\vec{r} - \vec{r'}}} \,.
 }
An explicit computation shows that $\left(-\nabla_{S^3}^2 + 3/4\right)  \langle z_a(r) z_b(r') \rangle_\text{free}^{S^3} = \delta_{ab} \delta^{(3)}(\vec{r} - \vec{r'}) / \sqrt{g(r)}$, which is indeed the equation of motion that would follow from \eqref{ZScalars} with $s = 3/4$ and $u = 0$.  This result was of course to be expected because a mass squared given by $s = 3/4$ corresponds to a conformally coupled scalar.

A more subtle issue is how to map to $S^3$ the interacting fixed point, which in flat space had $s = 0$ and $u = \infty$.  As explained for example in \cite{Gubser:2002vv}, the generating functional of connected correlation functions of the singlet operator $|z_a|^2$ in the theory with $u = \infty$ equals the Legendre transform of the corresponding generating functional in the theory with $u = 0$, to leading order in a large $N_b$ expansion.  This result holds on any manifold, and in particular on both $\R^3$ and $S^3$, and it assumes the other couplings in the theory are held fixed.  If we set $s =0$ on $\R^3$ and $s = 3/4$ on $S^3$, the Legendre transform assures us, for example, that to leading order in $N_b$ the two-point correlators in the theory with $u = \infty$ are
 \es{CorrelsInf}{
  \langle | z_a(r)|^2 |z_b(r')|^2 \rangle_{\text{critical}}^{\R^3} &= \frac{c N_b}{\abs{\vec{r} - \vec{r}'}^4} \,, \\
   \langle | z_a(r)|^2 |z_b(r')|^2 \rangle_{\text{critical}}^{S^3} &= \frac{c N_b  \left(1 + \abs{\vec{r}}^2 \right)^2 \left(1 + \abs{\vec{r}'}^2 \right)^2}{16\abs{\vec{r} - \vec{r}'}^4} \,,
 }
with the same normalization constant $c$, which is consistent with the conformal mapping of correlators realized through eq.~\eqref{Replacement}.   While in the free theory $z_a$ is a free field and the operator $|z_a|^2$  therefore has dimension $1$, in the interacting theory $|z_a|^2$ is a dimension $2$ operator. To study the interacting fixed point on $S^3$ we therefore should set $s = 3/4 + O(1/N_b)$ and take $u \to \infty$ in \eqref{ZScalarsEquiv}.

Reintroducing the fermionic and gauge fields,  we can write down the action as
 \es{LagRewrite}{
   S &= \int d^3r  \, \sqrt{g} \left[ \overline{\psi}_\alpha \gamma^\mu D_\mu \psi_\alpha + |D_\mu z_a|^2 + \left( s - i \lambda \right)  |z_a|^2 + \frac {1}{2u} \lambda^2  \right]
  + \frac{ik}{4 \pi} \int A \wedge dA \,.
 }
This action is of course invariant under gauge transformations, and therefore a correct definition of the path integral requires gauge fixing:
 \es{ZGaugeFixed}{
  Z = \frac{1}{\Vol(G)}  \int DA\, DX \, e^{-S[A, X]} \,,
 }
where $\Vol(G)$ is the volume of the group of gauge transformations, and $X$ denotes generically all fields besides the gauge field.
One justification for this normalization of the path integral is that for a pure Chern-Simons gauge theory on $S^3$
it yields the expected answer \cite{Witten:1988hf} $Z = 1/\sqrt{k}$, as will emerge from our computations below.
Because the first cohomology of $S^3$ is trivial, we can write uniquely any gauge field configuration $A$ as $A = B + d\phi$, where $d* B = 0$ and $\phi$ is defined only up to constant shifts.  One should think of $B$ as the gauge-fixed version of $A$ and of $d\phi$ as the possible gauge transformations of $A$.  Since the action $S[A, X]$ is gauge-invariant, it is independent of $\phi$ and only depends on $B$:  $S[A, X] = S[B, X]$.

We claim that
 \es{ChangeOfVars}{
  DA = DB\, D(d\phi) = DB\, D'\phi \sqrt{\det{}'\left(-\nabla^2\right)} \,,
 }
where $D'\phi$ means that we're not integrating over configurations with $\phi = \text{constant}$, and $\det'$ denotes the determinant with the zero modes removed from the spectrum.  To understand this relation, first note that the space $\Omega^p(S^3)$ of $p$-forms on $S^3$ is a metric space with the distance function ${\cal D}(\omega, \omega + \delta\omega) = \left( \int \abs{\delta \omega}^2 \right)^{1/2}$.  Then ${\cal D}(d\phi, d\phi + d \delta \phi) = \left( \int \abs{d\delta \phi}^2 \right)^{1/2}= \left( \int \delta\phi \left(-\nabla^2\right) \delta\phi \right)^{1/2}$ after integration by parts, and also ${\cal D}(\phi, \phi + \delta\phi) = \left( \int \abs{\delta \phi}^2 \right)^{1/2}$.  In other words, for each component of $\phi$ in a basis of eigenfunctions of the Laplacian, the distance between $d \phi$ and $d\phi + d \delta \phi$ is larger than the distance between $\phi$ and $\phi + \delta \phi$ by a factor of the square root of the eigenvalue with respect to $-\nabla^2$.  Eq.~\eqref{ChangeOfVars} follows as a straightforward change of variables.

The gauge transformations are maps from $S^3$ into the Lie algebra of the gauge group.  The volume of the group of gauge transformations $\Vol(G)$ can be expressed as
 \es{VolG}{
  \Vol(G) = \Vol(H) \int D'\phi \,,
 }
where $H$ is the group of constant gauge transformations, and $\int D'\phi$ is an integral over the non-constant gauge transformations with the measure given by the metric function ${\cal D}$ introduced in the previous paragraph.  In the case of a compact $U(1)$ with $\Vol(U(1)) = 2 \pi$, a constant gauge transformation $\phi = c$ has $c \in [0, 2 \pi)$.  Therefore
 \es{GotVolH}{
  \Vol(H) = \int_0^{2 \pi} dc \frac{{\cal D}(c, c + \delta c)}{\delta c} = \int_0^{2 \pi} dc\, \sqrt{\int 1} = 2 \pi \sqrt{\Vol(S^3)} \,.
 }
Combining \eqref{ZGaugeFixed}--\eqref{GotVolH} we obtain
 \es{PartFixed}{
  Z &=  \frac{{\cal C} \sqrt{\det{}'\left(-\nabla^2\right)}}{2 \pi \sqrt{\Vol(S^3)}} \int D\psi_\alpha\, Dz_a\, DB\, D\lambda \, e^{-S[\psi_\alpha, z_a, B, \lambda]}  \,.
 }
In this paper we will use the partition function in eq.~\eqref{PartFixed} to compute $F = -\log \abs{Z}$ in the limit where $N_f$, $N_b$, and $k$ are taken to be large and of the same order.

To leading order in the number of flavors we can ignore the gauge field and the Lagrange multiplier field $\lambda$.  Setting $s = 3/4$ as discussed above, we can write down the resulting path integral as
 \es{Z0}{
  Z_0 =  \int D\psi_\alpha\, Dz_a\, \exp\left[-\int d^3r  \, \sqrt{g} \left( \overline{\psi}_\alpha \gamma^\mu \nabla_\mu \psi_\alpha + |\partial_\mu z_a|^2 + \frac 34   |z_a|^2  \right) \right]  \,.
 }
In this approximation we have a theory of free $N_f$ Dirac fermions and $N_b$ complex scalars with the free energy \cite{Klebanov:2011gs}
 \es{Free0}{
   F_0  = \frac{\log 2}{4}\left( N_f  + N_b \right) + \frac{3\zeta(3)}{8 \pi^2} (N_f - N_b) \,.
 }

To find the corrections to $F_0$ we write \eqref{PartFixed} approximately as
 \es{PartFixedAgain}{
  Z \approx  e^{-F_0}  \frac{{\cal C}\sqrt{\det{}'\left(-\nabla^2\right)}}{2 \pi \sqrt{\Vol(S^3)}} \int DB\, D\lambda \, e^{- S_\text{eff}[\lambda] - S_\text{eff}^\text{vec}[B]} \,,
 }
with
 \es{Seff}{
  S_\text{eff}[\lambda] &= \int d^3 r \, \sqrt{g(r)} \frac {1}{2u} \lambda(r)^2
   - \frac 12 \int d^3 r\, \sqrt{g(r)} \int d^3 r'\, \sqrt{g(r')}  \lambda(r) \lambda(r') \left \langle |z_a(r)|^2 |z_b(r')|^2 \right \rangle_\text{free}^{S^3} \\
   S_\text{eff}^\text{vec}[B] &= \frac{ik}{4 \pi} \int B \wedge dB
     - \frac 12 \int d^3 r\, \sqrt{g(r)} \int d^3 r'\, \sqrt{g(r')}  B_\mu(r) B_\nu(r') \left \langle J^\mu(r) J^\nu(r') \right \rangle_\text{free}^{S^3} \,,
 }
where
 \es{JmuDef}{
  J^\mu(r) = \bar \psi_\alpha(r) \gamma^\mu \psi_\alpha(r) + i \bar z_a(r) \partial^\mu z_a(r) - i z_a(r) \partial^\mu \bar z_a(r)   \,.
 }
In writing the effective action \eqref{Seff} we used $\langle |z_a(r)|^2 \rangle_\text{free}^{S^3} = 0$, which follows from the fact that the free theory \eqref{Z0} is a CFT\@.

Defining
 \es{deltaFDefs}{
  \delta F_\lambda &= -\log \abs{ {\cal C} \int D\lambda\, e^{-S_\text{eff}[\lambda]} } \,, \\
  \delta F_A &= -\log \abs{  \frac{\sqrt{\det{}'(-\Delta)}}{2 \pi \sqrt{\Vol(S^3)}} \int DB\, e^{-S_\text{eff}^\text{vec}[B]}  } \,,
 }
we can then write $F = F_0 + \delta F_\lambda + \delta F_A + o(N^0)$.  The quantity $\delta F_\lambda$ was computed in \cite{Klebanov:2011gs}:
 \es{deltaFLambda}{
  \delta F_\lambda = -\frac{\zeta(3)}{8 \pi^2} \,.
 }
We devote the next section of this paper to calculating $\delta F_A$.

\section{Gauge field contribution to the free energy} \label{nonSUSY}

\subsection{Performing the Gaussian integrals}

Let's denote by $K^{\mu\nu}$ the integration kernel appearing in $S_\text{eff}^A$, namely
 \es{KmunuDef}{
  K^{\mu\nu}(r, r') &= -\langle J^\mu(r) J^\nu(r') \rangle_\text{free}^{S^3}  +  \frac{i k}{2 \pi} \frac{\delta^3(r-r')}{\sqrt{g(r)}} \frac{1}{\sqrt{g(r')}} \epsilon^{\mu \nu \rho} \partial_\rho'\,.
 }
As discussed above, when one writes $A = d \phi + B$ the action should be independent of $\phi$, so pure gauge configurations $A_\nu(r') = \partial_\nu' \phi(r')$ are exact zero modes of the kernel $K^{\mu\nu}(r, r')$.  Since we should integrate over the gauge-fixed field $B$ only, the Gaussian integration of the effective theory $S_\text{eff}^\text{vec}[B]$ yields $1/\sqrt{\det{}' (K^{\mu\nu} / (2 \pi))}$.  Again, the prime means that we remove the zero modes from the spectrum when we calculate the determinant.

Reinstating the radius $R$ of the sphere measured in units of some fixed UV cutoff, the discussion in the previous two paragraphs can be summarized as
 \es{FreeDifference}{
  \delta F_A = \frac{1}{2} \tr' \log \left[\frac{K^{\mu\nu}}{2 \pi R } \right] - \frac12 \tr' \log \left[-\frac{\nabla^2}{R^{2}}\right] +\log \left(2 \pi \sqrt{R^3 \Vol(S^3)} \right) \,,
 }
where all the operators are defined on an $S^3$ of unit radius.  Out of the first two terms in this expression, the second one is the easier one to calculate (see also \cite{Marino:2011nm}).  The spectrum of the Laplacian on a unit-radius $S^3$ consists of eigenvalues $n(n+2)$ with multiplicity $(n+1)^2$ for any $n \geq 0$.  One first rearranges the terms in the sum as
\es{Massless}{
\frac12 \tr' \log \left(-\frac{\nabla^2}{R^2}\right)&= \frac 12 \sum_{n=1}^\infty (n+1)^2 \log\frac{n(n+2)}{R^2}
    =\sum_{n=1}^\infty (n^2 + 1) \log \frac{n}{R}  - \frac{\log (2/R)}{2} \,.
 }
Then, using zeta-function regularization one writes
 \es{MasslessAgain}{
     \frac12 \tr' \log \left(-\frac{\nabla^2}{R^2}\right)  = - \frac{\log (2/R)}{2} - \frac{d}{ds}  \sum_{n=1}^\infty \frac{n^2 + 1}{(n/R)^s} \Biggr \rvert_{s=0}
    =  \frac{\zeta(3)}{4 \pi^2} + \frac{ \log (\pi R^2)}2  \,.
 }
Combining this expression with eq.~\eqref{FreeDifference} and using $\Vol(S^3) = 2 \pi^2$, we obtain
 \es{FA}{
  \delta F_A = \frac{1}{2} \tr' \log \left[\frac{K^{\mu\nu}}{2 \pi R } \right] - \frac{\zeta(3)}{4 \pi^2} +  \frac{\log  \left(8 \pi^3 R \right)}{2} \,.
 }
The only remaining task is the computation of the first term in this equation that we perform in the next subsection by explicit diagonalization of $K^{\mu\nu}$.

\subsection{Diagonalizing the kernel $K^{\mu \nu}(r,r')$}

Ultimately we would like to diagonalize the kernel $K^{\mu \nu}$ on $S^3$.  However, as a warm up it is instructive to consider the same diagonalization problem in flat space first.

 \subsubsection{Warm-up:  Diagonalization on $\R^3$}

The first step is to calculate the two-point function of current  $\langle J^\mu(r) J^\nu(0) \rangle_\text{free}^{\R^3} $, where we use the superscript $\R^3$ to emphasize that we are in flat space.  If we normalize the two-point functions of $z_a$ and $\psi_\alpha$ to be
 \es{CorrelpsiFlat}{
  \langle \bar z_a(r) z_b(0) \rangle_\text{free}^{\R^3} &= \int \frac{d^3 p}{(2 \pi)^3} \frac{\delta_{ab}}{\abs{p}^2} e^{i p \cdot r} = \frac{\delta_{ab}}{4 \pi \abs{r}} \\
  \langle \psi_\alpha(r) \bar \psi_\beta(0) \rangle_\text{free}^{\R^3}
   &= \int \frac{d^3 p}{(2 \pi)^3} \frac{\delta_{\alpha \beta} \gamma^\mu p_\mu}{\abs{p}^2} e^{i p \cdot r} = \frac{i}{4 \pi} \frac{\delta_{\alpha \beta} \gamma^\mu r_\mu }{\abs{r}^3} \,,
 }
then the two-point function of the current may be straightforwardly calculated to be
 \es{CorrelJFlat}{
  \langle J^\mu(r) J^\nu(0) \rangle_\text{free}^{\R^3} = \frac{N_f + N_b}{8 \pi^2} \frac{\abs{r}^2 \delta^{\mu\nu} - 2 r^\mu r^\nu}{\abs{r}^6} \,.
 }
It is simple to check that eq.~\eqref{CorrelJFlat} is of the right form.  This correlator is fixed up to an overall constant by the requirements that it should be homogeneous of degree $-4$ in $r$ ($J^\mu$ is a dimension 2 operator) and that away from $r = 0$ it should satisfy the conservation equation $\partial_\mu \langle J^\mu(r) J^\nu(0) \rangle_{\R^3}  = 0$.  Using
 \es{FTs}{
  \int d^3 r\, \frac{e^{i p \cdot r}}{\abs{r}^4} = -\pi^2 \abs{p} \,, \qquad
   \int d^3 r\, \frac{e^{i p \cdot r}}{\abs{r}^6} = \frac{\pi^2}{12} \abs{p}^3 \,,
 }
and introducing the Fourier space representation of the kernel $K^{\mu\nu}$ via
 \es{KFourier}{
  K_{\mu\nu}(r, r') = \int \frac{d^3 p}{(2 \pi)^3} K_{\mu\nu}(p) e^{i p\cdot (r - r')} \,,
 }
one obtains \cite{Kaul:2008xw}
\es{KmunuDefa}{
  K_{\mu\nu}(p) &= \frac{N_f + N_b}{16} \abs{p} \left(\delta_{\mu\nu} - \frac{p_\mu p_\nu}{\abs{p}^2} \right)
   +  \frac{k}{2 \pi} \epsilon_{\mu \nu}{}^{\rho} p_\rho\,.
 }
For fixed $p$, the eigenvalues of this matrix are
 \es{EvaluesFlat}{
  0 \,, \qquad  \frac{\abs{p}}{2} \left(\frac{N_f + N_b}{8} \pm i \frac{k}{\pi} \right) \,.
 }
The eigenvector associated with the zero eigenvalue is as expected $i p_\nu e^{i p \cdot r'}$, corresponding to a gauge configuration $A_\nu = \partial_\nu \phi$. We will now see that on $S^3$, while the diagonalization of $K^{\mu\nu}$ is significantly more complicated, the answer is equally simple:  the magnitude of the momentum $p$ appearing in \eqref{EvaluesFlat} should be replaced by a positive integer label $n$.

\subsubsection{Diagonalization on $S^3$}
\label{DIAGS3}

When we work with vector fields on $S^3$ it is convenient to introduce the dreibein
 \es{DreiDef}{
  e^i(r) = \frac{2}{1 + \abs{r}^2} dr^i
 }
and work only with frame indices.  For example,
 \es{JFrame}{
  \langle J^i(r) J^j(r') \rangle_\text{free}^{S^3} = e^i_\mu(r) e^j_\nu(r') \langle J^\mu(r) J^\nu(r') \rangle_\text{free}^{S^3} \,.
 }
The frame indices $i$ and $j$ are raised and lowered with the flat metric, so there is no distinction between lower and upper frame indices in Euclidean signature.

Using the transformation of correlators under Weyl rescalings in eq.~\eqref{Replacement}, one can immediately write down the current two-point function on $S^3$:
\es{CorrelJSphere}{
 \langle J^i(r) J^j(0) \rangle_\text{free}^{S^3}  = \frac{N_f + N_b}{8 \pi^2} \frac{\left(1+ \abs{r}^2\right)^2}{2}
    \frac{\abs{r}^2 \delta^{ij} - 2 r^i r^j}{\abs{r}^6} \,.
 }
As in flat space, the form of this correlator is fixed by the requirement that away from $r = 0$ we must have $\nabla_i \langle J^i(r) J^j(0) \rangle  = 0$.

To understand the diagonalization of $K^{ij}$ on $S^3$ we need to know that the space of square-integrable one-forms on $S^3$, being a vector space acted on by the $SO(4)  \cong SU(2)_L \times SU(2)_R$ isometry group, decomposes into irreducible representations of $SO(4)$ as follows.  Any one-form $\omega$ can be written as a sum of a closed one-form and a co-closed one-form.  The closed one-forms on $S^3$ are of course cohomologous to zero, so they're also exact.  A basis for them therefore consists of the gradients of the usual spherical harmonics.  Like the spherical harmonics, they transform in irreps with $j_L = j_R$.  On the other hand, the co-closed one-forms transform in irreps with $j_L = j_R \pm 1$.  So an arbitrary one-form can be written as
 \es{omegaDecomp}{
  \omega_i(r) = \sum_{n, \ell, m} \left[\omega^S_{n \ell m} \mathbb{S}^{n\ell m}_i(r) +
   \omega^L_{n \ell m} \mathbb{V}^{n\ell m}_{L, i}(r) + \omega^R_{n \ell m} \mathbb{V}^{n\ell m}_{R, i}(r)\right] \,,
 }
where we denoted by $\mathbb{S}^{n\ell m}_i$ the closed component transforming in the irrep with $j_L = j_R = (n-1)/2$ and by $\mathbb{V}^{n\ell m}_{L, i}$ and $\mathbb{V}^{n\ell m}_{R, i}$ the co-closed components transforming in the irreps with $j_R = j_L + 1 = n/2$ and $j_L = j_R + 1 = n/2$, respectively.  All the harmonics appearing in \eqref{omegaDecomp} have $n\geq 2$.  For $ \mathbb{S}^{n\ell m}_i$ there are $n^2$ states in each irrep indexed by the integers $\ell$ and $m$ satisfying $ 0 \leq \ell <n$ and $ -\ell \leq m \leq \ell$.   For the other two classes of vector harmonics, we have the same bounds on $m$ but now $ 0 < \ell <n$, giving a total dimension of $n^2 - 1$ for each irrep.

The $SO(4)$ generators commute with the kernel $K_{ij}$, so the eigenvectors of this kernel can be taken to be $\mathbb{S}^{n\ell m}_i$, $\mathbb{V}^{n\ell m}_{L, i}$, and $\mathbb{V}^{n\ell m}_{R, i}$.  The spectral decomposition of $K_{ij}$ is therefore
 \es{KSpectral}{
  K_{ij}(r, r') = \sum_{n, \ell, m} \left[s_n \mathbb{S}^{n\ell m}_i(r)  \mathbb{S}^{n\ell m}_j(r')^*
   + v_n^L \mathbb{V}^{n\ell m}_{L, i}(r)  \mathbb{V}^{n\ell m}_{L, j}(r')^*
    + v_n^R \mathbb{V}^{n\ell m}_{R, i}(r)  \mathbb{V}^{n\ell m}_{R, j}(r')^* \right] \,,
 }
where $s_n$, $v_n^L$, and $v_n^R$ are the corresponding eigenvalues.  These eigenvalues are independent of $\ell$ and $m$ because for fixed $n$ one can change $\ell$ and $m$ by acting with the $SO(4)$ generators, which commute with $K_{ij}$.  The degeneracy of $s_n$ is $n^2$ and that of $v_n^L$ and $v_n^R$ is $n^2 - 1$, with $n \geq 2$.

Given $K_{ij}$ one can find its eigenvalues by taking inner products with the eigenvectors.  Using rotational invariance, one can actually set $r'=0$ after summing over $\ell$ and $m$.  For example,
 \es{seigenvalues}{
  s_n &= \frac{\Vol(S^3)}{n^2} \sum_{\ell=0}^{n-1} \sum_{m=-\ell}^\ell \int_{S^3} d^3 r\, \mathbb{S}_i^{n\ell m}(r)^* K_{ij}(r, 0) \mathbb{S}_j^{n \ell m}(0)  \,,
 }
where the $n^2$ in the denominator is the dimension of the $SO(4)$ irrep to which $\mathbb{S}_i^{n\ell m}$ belong.
We notice that only the harmonics with $\ell =1$ contribute, so
 \es{snAgain}{
  s_n =  \frac{\Vol(S^3)}{n^2} \sum_{m=-1}^1 \int_{S^3} d^3 r\, \mathbb{S}_i^{n1 m}(r)^* K_{ij}(r, 0) \mathbb{S}_j^{n1m}(0)  \,,
 }
with similar expressions for $v_n^L$ and $v_n^R$, the only difference being that $n^2$ should be replaced by $n^2 - 1$.

Using explicit formulae for the harmonics (see Appendix~\ref{vecHarm}), one obtains
 \es{Gotsn}{
  s_n = \frac{N_f + N_b}{64 \pi n (n^2 - 1)} \int_0^\pi d\chi\, \csc^6 \frac{\chi}{2} \sin \chi
    \left[-2 n \cos(n \chi) \sin \chi + \left(1 - n^2 + \cos \chi( n^2 + 1)  \right) \sin (n \chi) \right]
 }
and
 \es{Gotvn}{
  v_n^{L, R} = \frac{N_f + N_b}{64 \pi n (n^2 - 1)} \int_0^\pi d \chi\, \csc^6 \frac{\chi}{2} \sin \chi
   \left[n \sin \chi \cos (n \chi) + \left(n^2 - n^2 \cos \chi - 1 \right) \sin n \chi \right]
    \pm \frac{i k n}{2 \pi} \,.
 }
The integration variable $\chi$ appearing here is related to $r$ through $\abs{r} = \cot (\chi/2)$.

We expect $s_n = 0$ because of gauge invariance.   Both \eqref{Gotvn} and \eqref{Gotsn} are divergent at $\chi = 0$, and need to be regulated.  A way of regulating them that doesn't preserve gauge invariance is to replace $\csc^6 \frac{\chi}{2}$ by $\csc^\alpha \frac{\chi}{2}$, compute these integrals for values of $\alpha$ for which they are convergent, and then formally set $\alpha = 6$.  Another way would be to assume $s_2 = 0$, and calculate $s_n - s_2$ and $v_n - s_2$, which are now convergent integrals.  Both of these ways of regulating \eqref{Gotsn} and \eqref{Gotvn} give
 \es{Gotvna}{
  s_n = 0 \,, \qquad v_n^L = \frac{n (N_f + N_b)}{16} +  \frac{i k n}{2 \pi} \,, \qquad
   v_n^R = \frac{n (N_f + N_b)}{16} -  \frac{i k n}{2 \pi} \,.
 }
Note the similarity between these expressions and the corresponding flat-space ones in eq.~\eqref{EvaluesFlat}.

\subsection{Contribution to the free energy}

We can now evaluate the first term in \eqref{FA}:
 \es{Free}{
  \frac{1}{2} \tr' \log \left(\frac{K^{\mu\nu}}{2 \pi R} \right)
    &= \sum_{n=2}^\infty (n^2 - 1) \log \abs{\frac{v_n^L}{2\pi R}}  \\
  &= \frac{1}{2} \log\left[ \frac{1}{8 \pi^2}
    \sqrt{ \left(\frac{N_f + N_b}{8}\right)^2 + \left( \frac{k}{\pi} \right)^2} \right]
     + \frac{\zeta(3)}{4 \pi^2}  - \frac{\log R}{2} \,,
  }
where the second line was obtained with the help of zeta-function regularization.  Combining this expression with \eqref{FA} yields
\es{Ffinal}{
\delta F_A = \frac{1}{2} \log\left[ \pi \sqrt{ \left(\frac{N_f + N_b}{8}\right)^2 + \left( \frac{k}{\pi} \right)^2} \right]  \,.
}
Note that all of the $\log R$ terms cancel in the final answer, as they should since we are describing a conformal fixed point, for which the path integral should be independent of $R$.   Another check of this result is that when $N_f  = N_b = 0$ we recover the standard result for the free energy of $U(1)$ CS theory on $S^3$~\cite{Witten:1988hf}, $\delta F_A = \frac 12 \log k$.

As an aside, we note that if we included the Maxwell term in \eqref{LA}, eq.~\eqref{Free} would be modified to
 \es{FreeModified}{
   \frac{1}{2} \tr' \log \left(\frac{K^{\mu\nu}}{2 \pi R} \right)
    &= \sum_{n=2}^\infty (n^2 - 1) \log \abs{\frac{1}{2 \pi R} \left( v_n^L + \frac{n^2}{e^2 R} \right)} \,,
 }
with $v_n^L$ still defined as in \eqref{Gotvna}.  Of course, $e^2$ flows to infinity in the IR, so as long as we have a non-zero CS level or non-zero numbers of flavors one can safely ignore the contribution from the Maxwell term in \eqref{FreeModified}.  If however one studies pure Maxwell theory with $k = N_f = N_b = 0$ so that $v_n^L = 0$ in \eqref{FreeModified}, the $S^3$ free energy becomes\footnote{We thank D. Jafferis and Z. Komargodski for very useful discussions of the free Maxwell field on $S^3$.}
\es{Fmax}{
F_{\text{Maxwell} } = -\frac{\log (e^2 R)}{2} + \frac{\zeta(3)}{4 \pi^2}   \,.
}
The logarithmic dependence on $R$ is consistent with the fact that the free Maxwell theory is not conformal in three spacetime dimensions.
$F_{\text{Maxwell} }$ decreases monotonically from the UV (small $R$) to the IR (large $R$).

\subsection{Generalization to $U(N_c)$ theory}

Eq.~\eqref{FA} generalizes straightforwardly to the case of $U(N_c)$ gauge theory with $N_f$ Dirac fermions and $N_b$ complex bosons transforming in the fundamental representation of the gauge group.  At large $k$ the contribution of the $\Tr A^3$ term in the Chern-Simons Lagrangian \eqref{LA} to the $S^3$ partition function is suppressed by $1/\sqrt{k}$, and the quadratic term proportional to $\Tr A \wedge  dA$ is the same as that of $N_c^2$ $U(1)$ gauge fields with Chern-Simons coupling $k$.  There are $N_f$ Dirac fermions and $N_b$ complex bosons charged under each of these $U(1)$ gauge fields.  One then just has to multiply the $U(1)$ answer \eqref{Ffinal} by a factor of $N_c^2$:
 \es{FColors}{
  \delta F_A = \frac{N_c^2}{2} \log\left[ \pi \sqrt{ \left(\frac{N_f + N_b}{8}\right)^2 + \left( \frac{k}{\pi} \right)^2} \right]
   + \log \frac{\Vol(U(N_c))}{\Vol(U(1))^{N_c^2}} \,.
 }
The second term in this expression comes from the different gauge fixing of the $U(N_c)$ gauge theory compared to a theory of $N_c^2$ $U(1)$ gauge fields.  As explained in section~\ref{largeNf}, the gauge fixing procedure involves dividing the partition function by the volume of the gauge group, so the partition function for the $U(N_c)$ theory has a prefactor of $1/\Vol(U(N_c))$ while the $U(1)^{N_c^2}$ theory obtained by multiplying \eqref{Ffinal} by $N_c^2$ would have a prefactor of $1/\Vol(U(1))^{N_c^2}$.  We have (see for example \cite{Marino:2011nm})
 \es{VolRatio}{
  \frac{\Vol(U(N_c))}{\Vol(U(1))^{N_c^2}} =
   \frac{(2 \pi)^{- N_c (N_c - 1) / 2}}{1! \cdot 2! \cdots (N_c - 1)!} \,.
 }
Thus, for $U(N_c)$ gauge theory with $N_f$ fundamental fermions and $N_b$ fundamental bosons we have
 \es{FUN}{
  F  = \frac{N_c \log 2}{4}\left( N_f  + N_b \right) + \frac{3\zeta(3)}{8 \pi^2} N_c (N_f - N_b)+
\frac{N_c^2}{2} \log\left[ \pi \sqrt{ \left(\frac{N_f + N_b}{8}\right)^2 + \left( \frac{k}{\pi} \right)^2} \right] \\
 - \frac 12 N_c (N_c - 1) \log (2 \pi) - \log \left(1! \cdot 2! \cdots (N_c - 1)! \right) + \ldots \,,
 }
with corrections expected to vanish in the limit of large $N_F$.  In writing \eqref{FUN} we kept $N_c$ of order one while scaling $N_f$, $N_b$, and $k$ to infinity with their ratios fixed.
Generalizing \eqref{FUN} to different gauge groups proceeds in a similar way.

\section{SUSY gauge theory with flavors} \label{SUSYscn}

In this section we compute the free energy of $U(1)$ Chern-Simons matter theories with ${\cal N} \geq 2$ supersymmetry coupled to a large number of flavors.  These computations allow us to check the first sub-leading correction to the non-SUSY result in the equation~\eqref{Ffinal} in a different way.  The computations in this section have as starting point the results of refs.~\cite{Kapustin:2009kz, Jafferis:2010un}, which used the technique of supersymmetric localization to rewrite the $S^3$ partition function of theories with ${\cal N} \geq 2$ SUSY as finite-dimensional integrals.  Our computations also involve finding the scaling
dimensions of the gauge-invariant operators.

\subsection{${\cal N}=4$ theory}

\begin{figure}[htb]
\begin{center}
\leavevmode
\scalebox{1}{\includegraphics{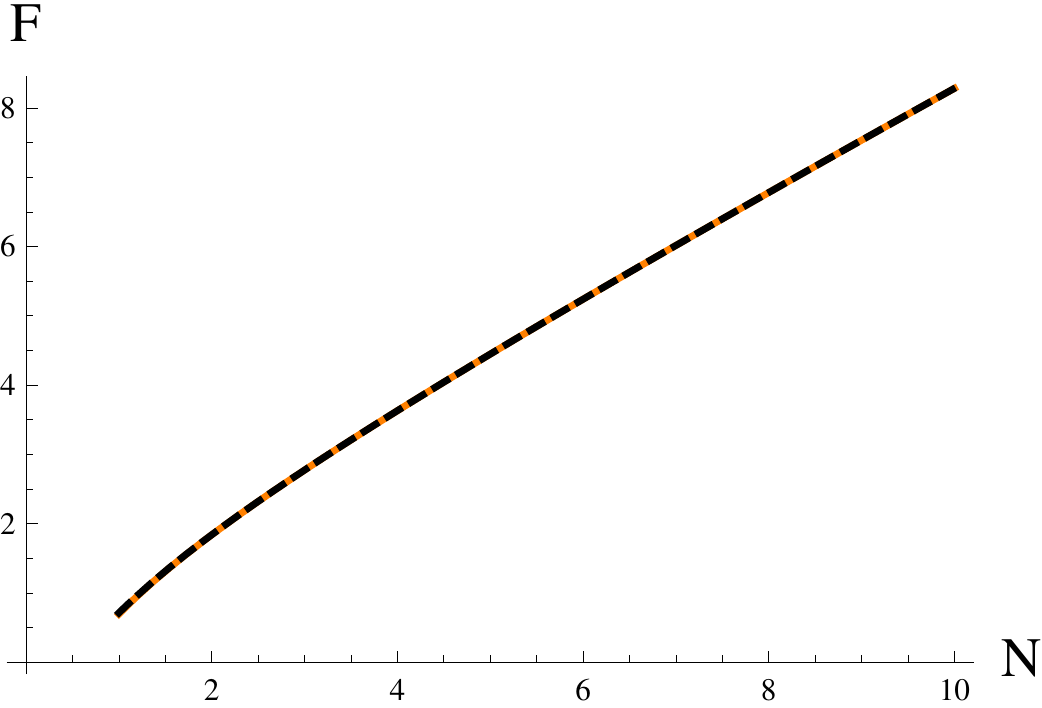}}
\end{center}
\caption{The exact free energy of the ${\cal N} =4$ theory obtained from eq.~\eqref{ZN4} (solid orange) and the analytical approximation \eqref{FN4} (dashed black).}
\label{PlotN4}
\end{figure}

As a warmup to the ${\cal N} = 2$ calculations, consider the ${\cal N } = 4$ parity-preserving supersymmetric $U(1)$ theory consisting of $N$ ${\cal N} = 4$ hypermultiplets coupled to an ${\cal N}= 4$ vector multiplet.   In ${\cal N} = 2$ notation, the ${\cal N}= 4$ vector multiplet consists of an ${\cal N} = 2$ vector and a neutral chiral superfield $\Phi$ of dimension $1$.  The ${\cal N}=4$ supersymmetry does not allow a Chern-Simons term. The hypermultiplets can be rewritten as $N$ pairs of oppositely charged chiral-multiplets $Q_a$ of $U(1)$ charge $+1$ and $\tilde Q_a$ with $U(1)$ charge $-1$.  The ${\cal N}=4$ SUSY requires a superpotential interaction $W \sim \tilde Q_a \Phi Q_a$.  The superpotential has R-charge equal to $2$. Then the $SU(2)$ subgroup of the $SO(4)_R$ R-symmetry, under which $\tilde Q_a$ and $\bar Q_a$ transform as a doublet, fixes the R-charge of the matter chiral multiplets to have the canonical free-field value:   $\Delta_Q = \Delta_{\tilde Q}= 1/2$.  The partition function is then given by~\cite{Kapustin:2009kz}
\es{ZN4}{
Z &= \frac{1}{2^{N}} \int_{-\infty}^\infty \frac{ d \lambda }{\cosh^{N}( \pi \lambda)} ={  2^{-N} \Gamma\left( {N \over 2} \right) \over \sqrt{\pi} \Gamma \left( {N + 1 \over 2} \right) } \,.
}
Expanding this at large $N$ we find
\es{FN4}{
F = -\log Z= N \log 2 + \frac12 \log\left( { N \pi \over 2} \right) - \frac{1}{4 N} + \frac{1}{24 N^3} + \ldots \,.
}
This large $N$ expansion is asymptotic, but it provides a very good approximation of the exact answer \eqref{ZN4} down to $N=1$---see Figure~\ref{PlotN4}.  Including more terms in \eqref{FN4} makes the approximation worse at $N=1$.

With $N$ pairs of  hyper-multiplets we have a total of $2 N$ physical complex bosons and $2 N$ Dirac fermions.  We then see perfect agreement of the first two terms in \eqref{FN4} with eqs.~\eqref{Free0} and~\eqref{Ffinal}.

\subsection{${\cal N}=3$ theory}

 Let us add the Chern-Simons term for the ${\cal N}=2$ abelian vector multiplet; it breaks ${\cal N}=4$ down to ${\cal N}=3$ supersymmetry.  The field content is the same as that of an ${\cal N} = 4$ vector multiplet and $N$ hypermultiplets, namely an ${\cal N} = 2$ vector, a neutral chiral $\Phi$, and $N$ pairs of chiral multiplets $Q_a$ and $\tilde Q_a$ charged under the ${\cal N} = 2$ vector.  The superpotential required by ${\cal N} = 3$ SUSY is
 \es{WN3}{
  W = -\frac{k}{4 \pi} \Phi^2 + \tilde Q_a \Phi Q_a \,.
 }
After integrating out the massive field $\Phi$, the superpotential can be rewritten as \cite{Gaiotto:2007qi}
 \es{WN3Again}{
  W = \frac{2 \pi}{k} (\tilde Q_a Q_a)^2 \,.
 }
The conformal dimensions of $Q_a$ and $\tilde Q_a$ are still equal to $1/2$, as is required by the marginality of $W$ and by the $\Z_2$ symmetry under which $Q_a$ and $\tilde Q_a$ are interchanged and all the fields in the vector multiplet change sign.  The partition function is \cite{Kapustin:2009kz}
\es{ZN3}{
Z &= \frac{1}{2^{N}} \int d \lambda \frac{e^{i \pi k \lambda^2}}{\cosh^{N}( \pi \lambda)} \,.
}
While this expression cannot be evaluated analytically, one can evaluate it using a saddle point approximation in the limit where both $k$ and $N$ are taken to be large.  Let us define $\kappa = 2 k / (N \pi)$ and take $N$ to infinity while keeping $\kappa$ fixed.  The saddle point is at $\lambda = 0$, and in order to obtain a systematic expansion, one should write
 \es{ZApproxN3}{
  Z = \frac{1}{2^{N}} \int d \lambda\, e^{-N  \pi^2 \lambda^2 \left(1 - i \kappa \right)/2}
   \left[1 + \frac{N \pi^4 \lambda^4}{12} -\frac{N \pi^6 \lambda^6}{45}
    + \frac{N(68 + 35 N) \pi^8 \lambda^8}{10080} +  \ldots  \right]\,,
 }
where the parenthesis contains the small $\lambda$ expansion of the function $e^{N \pi^2 \lambda^2 /2 } \cosh^{-N}(\pi \lambda)$.  Order by order in this expansion one can perform the integrals in \eqref{ZApproxN3} analytically.  The result is
 \es{ZApproxN3Explicit}{
  Z = \frac{1}{2^N} \sqrt{\frac{2}{N \pi (1 - i \kappa) }}
   \left[ 1 + \frac{1}{4 N (1 - i \kappa)^2}
    - \frac{1}{ 3 N^2 (1 - i \kappa)^3}
     + \frac{68 + 35 N}{96 N^3 (1- i \kappa)^4} +  \ldots \right] \,.
 }
Calculating $F = - \log \abs{Z}$ and expanding in $N$, we obtain
 \es{FApproxN3}{
  F = N \log 2 + \frac 12 \log \left(\frac{N \pi}{2} \sqrt{1 + \kappa^2} \right)
   + \frac{\kappa^2 - 1}{4(\kappa^2 + 1)^2 N}
    - \frac{4 \kappa^2(\kappa^2 - 1)}{3 (1 + \kappa^2)^4 N^2} + O(N^{-3}) \,.
 }
We note that in order to calculate the $O(N^{-\alpha})$ term in $F$ we need the expansion in \eqref{ZApproxN3} to be up to order $O(\lambda^{4\alpha})$.  The expression \eqref{FApproxN3} is also in agreement with eqs.~\eqref{Free0} and~\eqref{Ffinal} given that the ${\cal N}=3$ theory has $N_b = N_f = 2N$.

 Let us extend our discussion to the non-abelian theory with gauge group $U(N_c)$.  The field content now consists of an ${\cal N} = 4$ vector multiplet in the adjoint representation of the gauge group and $N$ pairs of chiral multiplets $Q_a$ and $\tilde Q_a$, in the fundamental and anti-fundamental representations of the gauge group, respectively.  After localization the partition function for this theory is given by~\cite{Kapustin:2009kz}
\es{ZN3Nc}{
Z &= \frac{2^{N_c(N_c-1)}}{2^{N N_c }N_c! }   \int \left( \prod_{i=1}^{ N_c}d \lambda_i \right) \left(\prod_{i < j}^{N_c}  \sinh^2[ \pi (\lambda_i - \lambda_j) ]  \right) \exp\left(i \pi k \sum_{i=1}^{N_c} \lambda_i^2\right) \prod_{i=1}^{N_c} \cosh^{-N} ( \pi \lambda_i) \,.
}
In the limit where $N_c / N \ll 1$, the integral has a saddle point at $\lambda_i =  O\left[ (N_c / N)^{1/2} \right]$.  Through next to leading order the partition function of the non-abelian theory reduces to
\es{ZNc2}{
Z &= \frac{(2 \pi)^{N_c (N_c  - 1)}}{2^{N N_c} N_c! N^{N_c^2 \over 2}} \left( \int \prod_{i=1}^{ N_c}d \tilde \lambda_i \right)  \left(\prod_{i < j}^{N_c} (\tilde \lambda_i - \tilde \lambda_j)^2 \right) \exp\left(- {\pi^2 \left(1 - i \kappa \right) \over 2} \sum_{i=1}^{N_c} \tilde \lambda_i^2 \right) + \ldots \,,
}
where $\kappa$ is defined as in the abelian theory and $\tilde \lambda_i = \sqrt{N} \lambda_i$.  We have rescaled the integration variables so that the remaining integrals in eq.~\eqref{ZNc2} produce numbers independent of $N$.  Taking the log of eq.~\eqref{ZNc2} we then see immediately that
\es{Freec2}{
F = N_c N \log 2 + \frac{N_c^2}{2} \log(N) + O(N^0) \,.
}
Given that the non-abelian theory has $N_b = N_f = 2 N$, the equation above is in agreement with eq.~\eqref{FUN}.

\subsection{Non-chiral ${\cal N}=2$ theory}

Moving up one notch in complexity, we now consider the ${\cal N}=2$ Chern-Simons theory coupled to the chiral fields $Q_a$ and $\tilde Q_a$ introduced above, this time without the superpotential \eqref{WN3Again}.  The absence of the superpotential leaves the R-charges of $Q_a$ and $\tilde Q_a$ a priori unrestricted.  It was proposed in \cite{Jafferis:2010un} that one way of finding the correct IR R-charges in an ${\cal N}=2$ theory is by calculating the partition function on $S^3$ for any choice of trial R-charges consistent with the marginality of the superpotential and then extremizing over all such R-charge assignments.  The R-charges of $Q_a$ and $\tilde Q_a$ can be taken to be equal to some common value $\Delta$ because of the following symmetries:  the action is invariant under two $U(N)$ symmetries under which the $Q_a$ and $\tilde Q_a$ transform as fundamental vectors, as well as under a charge conjugation symmetry that flips the sign of all the fields in the vector multiplet and at the same time interchanges $Q_a$ and $\tilde Q_a$.

As a function of $\Delta$, the partition function is \cite{Jafferis:2010un}
\es{ZN2}{
Z = \int_{-\infty}^\infty d \lambda\, e^{i \pi k \lambda^2} e^{N \big (\ell(1 - \Delta + i \lambda) +  \ell(1- \Delta - i \lambda)\big ) } \,,
}
where the function $\ell(z)$ is given by
\es{lz}{
\ell(z) = -z \log\left( 1 - e^{2\pi i z} \right) + {i\over2} \left( \pi z^2 + \frac 1 \pi \text{Li}_2 \left( e^{2\pi i z} \right) \right) - \frac{i \pi}{12} \, .
}
This function can be found by solving the differential equation $\partial_z \ell(z) = - \pi z \cot( \pi z)$ with the boundary condition $\ell(0) = 0$.  It is a real function when $z$ is real.

We again take $N$ to infinity while keeping $\kappa = 2 k / (N \pi)$ fixed.  In this limit one can use the saddle point approximation to calculate the partition function \eqref{ZN2} as in the previous section.  The exponent in \eqref{ZN2} is an even function of $\lambda$, so there is a saddle point at $\lambda = 0$, and we will assume this is the only relevant saddle.  To leading order in $N$ we therefore have
 \es{FN2Leading}{
  F(\Delta) = -2 N \ell(1 - \Delta) + O(\log N)\,.
 }
This function is maximized when $dF / d\Delta = 2 \pi (\Delta - 1) \cot (\pi \Delta) = 0$, which implies $\Delta = 1/2 + O(N^{-1})$. We will find that this anomalous dimension affects $F$ only at order $1/N$, i.e. the first two leading orders in the large $N$ expansion of $F$ are the same for the ${\cal N}=2$ theory and the ${\cal N}=3$ theory studied in the previous section.

\begin{figure}[htb]
\begin{center}
\leavevmode
\scalebox{1.1}{\includegraphics{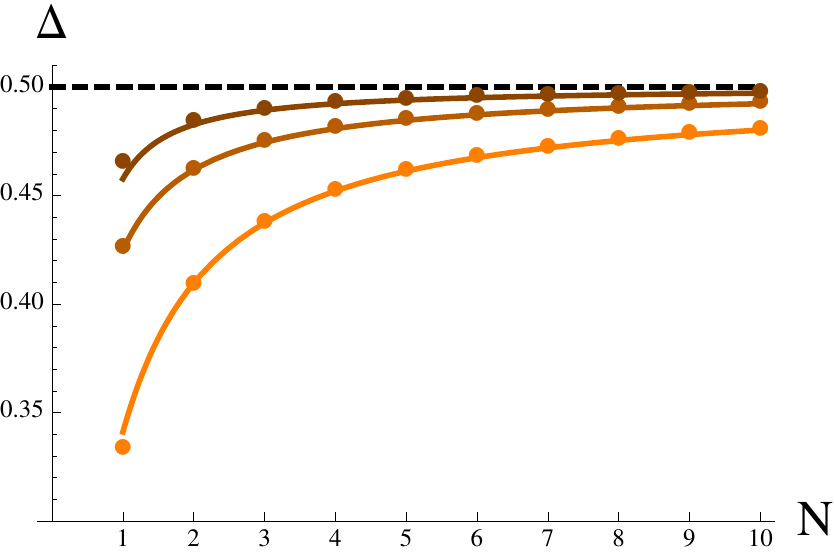}}
\end{center}
\caption{ The R-charge $\Delta$ plotted as a function of $N$ for $\kappa = 0, 4/\pi, 8/\pi$, with darker plots corresponding to larger $\kappa$.  The solid lines are calculated using the approximation in eq.~\eqref{DeltaN2}.   The circles are computed by numerically maximizing the free energy with respect to $\Delta$.  Note that the two computations match well even for small $N$. }
\label{Deltaplot}
\end{figure}

One can develop a systematic expansion to study $1/N$ corrections in a similar way to what was done at the end of the previous section for the ${\cal N} = 3$ theory.  The fact that now $\Delta$ depends on $N$ introduces an extra complication.  We expand $\Delta$ as
 \es{DeltaExpansion}{
  \Delta = \frac 12 + \frac{\Delta_1}{N} + \frac{\Delta_2}{N^2} + \ldots \,,
 }
and we rescale $\lambda = \tilde \lambda / \sqrt{N}$.  One can then write
\es{ZN2Approx}{
Z = \frac{1}{2^N \sqrt{ N}} \int_{-\infty}^\infty d \tilde \lambda\,
   e^{- \pi^2 \tilde \lambda^2 \left(1 - i \kappa \right)/2} \left[
   1 + \frac{6 \pi^2 \Delta_1^2 + 24 \Delta_1 \tilde \lambda^2 + \tilde \lambda^4}{12 N}
    + \ldots
   \right]
 \,,
}
where the expansion in parenthesis is in powers of $1/N$ while holding $\tilde \lambda$ fixed.  Term by term in this expansion, these integrals can be evaluated analytically.  The free energy is
 \es{FreeN2}{
   F(\Delta) = N \log 2 + \frac 12 \log \left(\frac{N \pi}{2} \sqrt{1 + \kappa^2} \right)
   - \left(\frac{\pi^2 \Delta_1^2}{2} + \frac{2 \Delta_1}{1 + \kappa^2} + \frac{1 - \kappa^2}{4 (1 + \kappa^2)^2} \right) \frac{1}{N} + \ldots \,.
 }
Maximizing this expression with respect to $\Delta_1$ we obtain
 \es{GotDelta1}{
  \Delta_1 = -\frac{2}{\pi^2 (1 + \kappa^2)} \,.
 }
For $k\gg N \gg 1$ this result agrees with section 6.3 of \cite{Jafferis:2010un}.
Repeating this procedure two more orders in $F$ we find
 \es{DeltaN2}{
  \Delta = \frac 12 - \frac{2}{\pi^2 (1 + \kappa^2)} \frac 1N - \frac{2 \left[\pi^2 - 12 + \kappa^2 (4 - 2 \pi^2) + \pi^2 \kappa^4 \right]}{\pi^4 (1 + \kappa^2)^3} \frac 1{N^2} + O(N^{-3}) \,.
 }
This series appears to be perfectly convergent. In fig.~\ref{Deltaplot} we plot $\Delta (N)$ for a few values of $\kappa$ using both the precise numerical result
and the approximation \eqref{DeltaN2}.

Using eqs.~\eqref{FreeN2} and \eqref{GotDelta1} we find that the free energy is
 \es{FreeN2Final}{
  F = N \log 2 + \frac 12 \log \left(\frac{N \pi}{2} \sqrt{1 + \kappa^2} \right)
   + \left( \frac{\kappa^2 - 1}{4 (1 + \kappa^2)^2} +  \frac{2}{\pi^2 (1 + \kappa^2)^2} \right) \frac{1}{N} + O(N^{-2}) \,.
 }
Using $N_b =N_f = 2N$, we see that this expression agrees with eqs.~\eqref{Free0} and~\eqref{Ffinal} that were derived directly from a large $N$ expansion without the use of supersymmetric localization.

Let us perturb the ${\cal N} =2$ theory discussed above by the quartic superpotential
 \es{WMoreGeneral}{
  W = g (Q_a \tilde Q_a)^2 \,.
 }
Since, as can be seen from \eqref{GotDelta1}, the dimension of $Q_a$ and $\tilde Q_a$ is slightly smaller than $1/2$, the perturbation \eqref{WMoreGeneral} is a slightly relevant perturbation of the UV ${\cal N} = 2$ theory.
This theory should flow to an IR fixed point where the superpotential is exactly marginal, i.e. the IR R-charges of $Q_a$ and $\tilde Q_a$
are $1/2$. The calculation of $F_\text{IR}$ is thus exactly the same as for the ${\cal N}=3$ superconformal $U(1)$ theory
discussed in section 4.2. The infrared theory is conformal for any $g$, and for the special value $g = 2 \pi/ k$ it is the ${\cal N} = 3$ theory in eq.~\eqref{WN3Again}.  Eqs.~\eqref{FApproxN3} and \eqref{FreeN2Final} imply that the change in free energy between the UV and IR fixed points is
 \es{DeltaF}{
  F_\text{UV} - F_\text{IR} =  \frac{2}{\pi^2 (1 + \kappa^2)^2 N}  + O(N^{-2}) \,,
 }
which can be explicitly seen to be positive, in agreement with the conjectured $F$-theorem \cite{Jafferis:2011zi}.

Since the superpotential deformation \eqref{WMoreGeneral} is only slightly relevant, one may wonder how the result \eqref{DeltaF} compares with the perturbative computation performed in \cite{Klebanov:2011gs}.  In \cite{Klebanov:2011gs} it was shown that if the Lagrangian is perturbed by a slightly relevant scalar operator of dimension $3 - \epsilon$, then there is a perturbative IR fixed point and $F_\text{UV} - F_\text{IR} \propto \epsilon^3$.  If however the Lagrangian is perturbed by a pseudoscalar operator of dimension $3 - \epsilon$, then there is no perturbative fixed point;  it was seen in an example that if a fixed point exists then one might expect $F_\text{UV} - F_\text{IR} \propto \epsilon$.  In our case, the superpotential deformation \eqref{WMoreGeneral} translates into perturbations of the Lagrangian by both a scalar operator ${\cal O}_1$ and a pseudoscalar operator ${\cal O}_2$.  Indeed, denoting
 \es{QExpansion}{
  Q_a = \phi_a + \sqrt{2} \theta \psi_a + \theta^2 F_a \,, \qquad
   \tilde Q_a = \tilde \phi_a + \sqrt{2} \theta \tilde \psi_a +\theta^2 \tilde F_a \,,
 }
we have
 \es{DeltaL}{
  \delta {\cal L} &=  g^2 {\cal O}_1 + g {\cal O}_2 \,, \\
   {\cal O}_1 &= - 8 \abs{\phi_a \tilde \phi_a \phi_b}^2
   - 8 \abs{\phi_a \tilde \phi_a \tilde \phi_b}^2 \,, \\
  {\cal O}_2 &= - 2 \psi_a \tilde \psi_a \phi_b \tilde \phi_b - \psi_a \psi_b \tilde \phi_a \tilde \phi_b - \tilde \psi_a \tilde \psi_b \phi_a \phi_b - 2 \psi_a \tilde \psi_b \tilde \phi_a \phi_b + \text{c.c} \,.
 }
The scaling dimensions of these operators are
 \es{ScalingDims}{
  \Delta({\cal O}_1) = 3 + 6 \frac{\Delta_1}{N} + O(N^{-2}) \,, \qquad
   \Delta({\cal O}_2) = 3 + 4 \frac{\Delta_1}{N} + O(N^{-2}) \,,
 }
so the pseudoscalar operator ${\cal O}_2$ is the more relevant one.  One might expect the IR fixed point should be non-perturbative and that $F_\text{UV} - F_\text{IR} \propto -\Delta_1/ N$ times a function of order one.  That the IR fixed point is non-perturbative can be seen after writing $g = \hat g / N$ so that $\hat g$ stays of order $1$ as we take $N$ to infinity.   The IR coupling $g_\text{IR} = 2 \pi / k$ corresponds to $\hat g_\text{IR} = 4 / \kappa$, which is of order one in the large $N$ limit, meaning that the IR fixed point is non-perturbative.  That $F_\text{UV} - F_\text{IR} \propto -\Delta_1/ N$ times a function of order one can be immediately seen from eqs.~\eqref{DeltaF} and \eqref{GotDelta1}.

\subsection{Chiral ${\cal N} = 2$ theory}

We now consider a natural generalization of the non-chiral ${\cal N} = 2$ theory discussed in the previous section---the chiral ${\cal N} = 2$ theory.  This theory is given by ${\cal N} = 2$ Chern-Simons theory coupled to $N$ chiral fields $Q_a$ and $\tilde N$ anti-chiral
fields $\tilde Q_a$ with no superpotential.  When $N = \tilde N$ this theory reduces to the non-chiral theory discussed in the previous section.  Without loss of generality, we now assume that $N > \tilde N$.  Instead of dealing with $N$ and $\tilde N$ it is convenient to define the following quantities,
\es{npm}{
 \bar N \equiv { N + \tilde N \over 2} \,, \qquad \mu \equiv { N - \tilde N \over N + \tilde N} \,, \qquad 0 < \mu \leq 1 \,.
 }

The R-charges of $Q_a$ and $\tilde Q_a$, which we denote by $\Delta$ and $\tilde \Delta$, are not gauge-invariant observables. 
Gauge invariant operators may be constructed from combinations of $Q_a$, $\tilde Q_a$, and the monopole operators $T_m$, which create $m$ units of magnetic flux through $2$-spheres surrounding their insertion points.  The R-charge of $T_m$ is given by~\cite{Benini:2009qs,Jafferis:2009th}
\es{Rtm}{
R[T_m] = \gamma_{ |m| } + m \, \delta \,,
}
where $ \gamma_{ |m| }$ is determined in terms of $\Delta$ and $\tilde \Delta$, while $\delta$ is so far arbitrary.
In the $F$-maximization procedure one finds that in the space of $\delta$, $\Delta$, and $\tilde \Delta$ there is exactly one flat direction:  $F$ remains unchanged if we send simultaneously $\Delta \to \Delta + r$, $\tilde \Delta \to \tilde \Delta - r$, and $\delta \to \delta + k r$ for any $r$ \cite{Jafferis:2011zi}.  The R-charges of the gauge-invariant operators are of course independent of $r$.  As long as $k \neq 0$, we can set $\delta = 0$ as a gauge choice and work only with $\Delta$ and $\tilde \Delta$, which are not necessarily equal when $\mu \neq 0$.

  As a function of the R-charges $\Delta$ and $\tilde \Delta$, the partition function we need to consider is then
\es{ZN2chiral}{
Z = \int_{-\infty}^\infty d \lambda\, e^{i \pi k \lambda^2} e^{N \ell(1 - \Delta + i \lambda) + \tilde N  \ell(1- \tilde \Delta - i \lambda) } \,.
}
We want to calculate the partition function in the limit where $\bar N$ goes to infinity and $\kappa = 2 k / (\bar N \pi)$ and $\mu$ are held fixed.
In the large $\bar N$ limit we again find a saddle point at $\lambda = 0$.  The saddle point equation requires $ \Delta = 1/2 + O(1/ \bar N)$ and $\tilde \Delta = 1/2 + O(1/ \bar N)$.  In order to study $1/\bar N$ corrections, we expand the R-charges as
 \es{DeltaExpansionchiral}{
  \Delta = \frac 12 + \frac{\Delta_1}{\bar N} + \frac{\Delta_2}{ \bar N^2} + \ldots \,, \qquad \tilde \Delta = \frac 12 + \frac{\tilde \Delta_1}{\bar N} + \frac{\tilde \Delta_2}{\bar N^2} + \ldots \,.
 }
Using the methods developed in the previous sections, we can calculate the free energy perturbatively in the $1/\bar N$ expansion and maximize the resulting expression term by term with respect to the $\Delta_i$ and $\tilde \Delta_i$.  Going through the procedure we find the following results for the free energy and the R-charges:
\es{resultsChiral}{
\Delta &= \frac12 - \frac{2 ( 1 + \mu)}{\pi^2 ( 1 + \kappa^2) \bar N} + O(\bar N^{-2}) \,, \qquad \tilde \Delta = \frac12 - \frac{2 ( 1 - \mu)}{\pi^2 ( 1 + \kappa^2) \bar N} + O(\bar N^{-2}) \,, \\
 F &= \bar N \log 2 + \frac 12 \log \left(\frac{\bar N \pi}{2} \sqrt{1 + \kappa^2} \right)
   + \left[ \frac{\kappa^2 - 1}{4 (1 + \kappa^2)^2} +  \frac{2}{\pi^2 (1 + \kappa^2)^2}  \right. \\
   &\left. - \frac{4 \mu^2}{ \pi^2} \left( \frac{1}{(1+\kappa^2)^2} - \frac{4}{3 (1+\kappa^2)^3} \right) \right] \frac{1}{\bar N} + O(\bar N^{-2}) \,.
}
Using $N_b =N_f = 2 \bar N$, we see that the expression for $F$ agrees with eqs.~\eqref{Free0} and~\eqref{Ffinal} that were derived without the use of supersymmetric localization.  The combination $\Delta + \tilde \Delta$, which gives the R-charge of the the gauge invariant meson operators $Q_a \tilde Q_b$, is in agreement with the pertrubative calculations in \cite{Amariti:2011da,Amariti:2011jp}.

We can perturb this theory by adding in the superpotential\footnote{Changing the relative coefficients of the terms in \eqref{chirlsuper} is an exactly marginal deformation \cite{Chang:2010sg} and does not change $F$.}
\es{chirlsuper}{
W \sim \sum_{a,b} (Q_a \tilde Q_b)^2 \,.
}
Since $\Delta + \tilde \Delta < 1$ in the UV ${\cal N} = 2$ CFT, this superpotential deformation is relevant and causes an RG flow to the fixed point where the superpotential is exactly marginal.  At the IR ${\cal N} = 2$ fixed point we have the constraint $\Delta +\tilde \Delta = 1$.  To determine the free energy at the IR fixed point we simply have to repeat the $F$-maximization procedure above subject to this constraint.  That in the UV one has to maximize $F$ without any constraints while in the IR one has to maximize $F$ under the constraint $\Delta +\tilde \Delta = 1$ means that the free energy of the IR fixed point is necessarily at most equal to the free energy of the UV fixed point.  Indeed, we find that
\es{resultsChiralIR}{
F_\text{UV} - F_\text{IR} = \frac{2 (1 - \mu^2) }{ \pi^2 (1 + \kappa^2)^2} \frac{1}{\bar N}  + O(\bar N^{-2}) \,,
}
which is manifestly positive when $\mu^2 < 1$.  When $\mu = 1$ there are no $\tilde Q_a$ fields, and so we are not allowed to add in the superpotential deformation.
The R-charges at the IR fixed point are given by
\es{RchargeIRchiral}{
\Delta = \frac12 - \frac{4 \mu}{ \pi^2 (1+ \kappa^2)} \frac{1}{\bar N} + O(\bar N^{-2})     \,, \qquad \tilde \Delta = 1 - \Delta \,.
}

\section{Discussion}

In this paper we have studied certain 3-dimensional gauge theories coupled to a large number $N_F$ of massless charged fields.
Such theories are conformal for a sufficiently large $N_F$, and a good tool for studying them is the $1/N_F$ expansion.
In this paper we used such an expansion to study the disk entanglement entropy, which is related to the free energy $F$ on the 3-sphere.

For the $U(N_c)$ gauge theory coupled to $N_f$ massless Dirac fermions and $N_b$ massless scalars we found the first subleading term in the
expansion, \eqref{FUN}. We have also studied the ${\cal N}=2$ supersymmetric abelian gauge theory coupled to $N$ positively charged chiral superfields $Q$ and $N$ negatively charged chiral superfields $\tilde Q$.  In this case, $F$ can be calculated numerically for any $N$ using the methods of localization. We compared these numerical results with their $1/N$ expansion and found excellent agreement down to small $N$.

An important question concerning such CFTs is whether there is a breakdown of conformal invariance for sufficiently small $N_F$.
In the ${\cal N}=2$ supersymmetric $U(1)$ gauge theory, even for a single non-chiral flavor the theory is conformal and unitary.
This is indicated by the mirror symmetry arguments~\cite{Aharony:1997bx} and confirmed by explicit calculation of the localized path integral in
\cite{Jafferis:2010un}, which indicates that the dimension of $Q$ and $\tilde Q$ is exactly $1/3$.
However, in the non-supersymmetric $U(1)$ theories there typically is a lower bound for the conformal window.
For example, in the extreme limit $N_F=0$ we find the free Maxwell theory, which is not conformally invariant. We studied it on the $S^3$ of radius $R$ in section 3 and found that $F_{\text{Maxwell}}$
varies logarithmically with $R$, eq. (\ref{Fmax}), indicating the lack of conformal invariance.

One possible phenomenon for small $N_F$ is the chiral symmetry breaking in 3-dimensional QED coupled to massless fermions \cite{Appelquist:1986fd,Appelquist:1988sr}.
The numerical studies of lattice antiferromagnets \cite{assaad} suggest
the QED theory with $N_f=8$ Dirac fermions is a stable CFT, while the $N_f=4$ theory is unstable to symmetry breaking towards a
non-conformal ground state \cite{hermele2}.
More generally, one of the signs of crossing the lower edge of the conformal window could be that the assumption of conformality leads to certain gauge invariant operators having scaling dimensions that violate the 3-dimensional unitarity bound $\Delta> 1/2$.

In \cite{Jafferis:2011zi,Klebanov:2011gs} it was conjectured that $F$ must be positive in a unitary CFT.
Since as $N_F$ decreases so does $F$, it is possible that $F$ may become negative for sufficiently small $N_F$. This could serve as
another criterion for theories outside the conformal window. It would be interesting to explore the different criteria above and to see
if they are related.

\section*{Acknowledgments}
We thank A.~Amariti, A.~Dymarsky,
D.~Jafferis, Z.~Komargodski, T.~Nishioka, N.~Seiberg, M.~Siani, and E.~Witten for helpful discussions.
 The work of IRK was supported in part by the US NSF under Grant No.~PHY-0756966. IRK gratefully acknowledges support from the IBM Einstein Fellowship at the Institute for Advanced Study, and from the John Simon Guggenheim Memorial Fellowship.  SSP was supported by a Pappalardo Fellowship in Physics at MIT and by the U.S. Department of Energy under cooperative research agreement Contract Number DE-FG02-05ER41360\@.  The work of SS was supported by the National Science Foundation under grant DMR-1103860 and by a MURI grant from AFOSR.  BRS was supported by the NSF Graduate Research Fellowship Program. BRS thanks the Institute for Advanced Study for hospitality.

\appendix

\section{Vector spherical harmonics on $S^3$}  \label{vecHarm}

In this appendix we review the vector spherical harmonics on $S^3$ (see for example \cite{Tomita:1982ew}).
The scalar spherical harmonics on $S^3$ transform in the $[j, j]$ irrep of $SU(2) \times SU(2)$ for any half-integer $j \geq 0$.  Denoting $n = 2j+1$, the dimension of this representation is $n^2$, and $n \geq 1$ is an integer.  Let's denote these harmonics by $Y_{n\ell m}$, with $0 \leq \ell < n$ and $-\ell \leq m \leq \ell$.  Parameterizing $S^3$ by three angles $(\chi, \theta, \phi)$, with the line element given by
 \es{S3Line}{
  ds^2 = d\chi^2 + \sin^2 \chi\, \left[ d\theta^2 + \sin^2\theta\, d\phi^2 \right] \,,
 }
one can write down explicit formulas for the spherical harmonics:
 \es{GotY}{
  Y_{n \ell m}(\chi, \theta, \phi) = \frac{\sin^\ell \chi}{\sqrt{a_{n \ell}}} \Phi_{n \ell}(\chi) Y_{\ell m}(\theta, \phi)\,,
 }
where $Y_{\ell m}$ are the usual spherical harmonics on $S^2$ and
 \es{aPhiDefs}{
  \Phi_{n \ell}(\chi) = \frac{d^{\ell + 1} \cos (n \chi)}{d(\cos \chi)^{\ell +1}} \,, \qquad
   a_{n \ell} = \frac{n \pi}{2} \frac{(\ell + n)!}{(n - \ell - 1)!} \,.
 }
The spherical harmonics satisfy
 \es{YEq}{
  \left( \nabla^\mu \nabla_\mu + n^2 - 1 \right) Y_{n \ell m} = 0 \,.
 }

As mentioned in the main text, any one-form on $S^3$ can be written as a linear combination of forms that transform in $SO(4)$ irreps with $j_L = j_R$, $j_L = j_R + 1$ and $j_R = j_L + 1$.  One way to understand this fact is as follows.  One can express any one-form as a linear combination of the right-invariant one-forms on $S^3$.  These right-invariant one-forms transform in the $[1, 0]$ irrep.  The coefficients of the right-invariant forms in the decomposition of an arbitrary form can in turn be expanded in terms of the usual spherical harmonics, which as mentioned above transform in the $[j, j]$ irreps.     The Hilbert space of square-integrable one-forms therefore decomposes as
 \es{ProdDecomp}{
  \bigoplus_j \left( [1, 0] \otimes [j, j] \right) = \bigoplus_j \left( [j-1, j] \oplus [j, j] \oplus [j+1, j] \right) \,,
 }
The sum runs over all $j \in \N /2$, but when $j = 1/2$ the first term in the paranthesis is absent, and when $j = 0$ the first two terms are absent.  Switching from $j$ to $n$ we see that the one-forms on $S^3$ transform in
 \es{VectorDecomp}{
   \bigoplus_{n = 2}^\infty \left[\frac{n-1}{2}, \frac{n-1}{2} \right]
   \oplus
    \bigoplus_{n=2}^\infty  \left[\frac{n-2}{2}, \frac{n}{2} \right]
     \oplus \bigoplus_{n=2}^\infty  \left[\frac{n}{2}, \frac{n-2}{2} \right]  \,.
 }
We will call the harmonics corresponding to the first sum $\mathbb{S}^{n \ell m}_\mu$ (with $0 \leq \ell < n$ and $-\ell \leq m \leq \ell$, having total dimension $n^2$), and those corresponding to the second and third sums $\mathbb{V}^{n \ell m}_{L, \mu}$ and $\mathbb{V}^{n \ell m}_{R,\mu}$, respectively (with $0 < \ell < n$ and $-\ell \leq m \leq \ell$, having total dimension $n^2-1$ for either $\mathbb{V}^{n \ell m}_{L, \mu}$ or $\mathbb{V}^{n \ell m}_{R,\mu}$).

Explicit formulas are available.  Defining the inner product on the space of square-integrable one-forms in the usual way,
 \es{Inner}{
  \langle \mathbb{A}, \mathbb{B} \rangle =
   \int d\chi\, d\theta\, d\phi\, \sin^2\chi \sin \theta\,\mathbb{A}_\mu(\chi, \theta, \phi)^*\, \mathbb{B}^\mu(\chi, \theta, \phi) \,,
 }
we normalize the harmonics so that
 \es{Normalizations}{
  \langle \mathbb{S}^{n\ell m}, \mathbb{S}^{n'\ell' m'}\rangle
   &= \langle \mathbb{V}^{n\ell m}_L, \mathbb{V}^{n'\ell' m'}_L\rangle
    = \langle \mathbb{V}^{n\ell m}_R, \mathbb{V}^{n'\ell' m'}_R\rangle
   = \delta_{nn'} \delta_{\ell \ell'} \delta_{mm'} \\
   \langle \mathbb{S}^{n\ell m}, \mathbb{V}^{n'\ell' m'}_L\rangle
    &= \langle \mathbb{S}^{n\ell m}, \mathbb{V}^{n'\ell' m'}_R\rangle
    = \langle \mathbb{V}^{n\ell m}_L, \mathbb{V}^{n'\ell' m'}_R\rangle  = 0 \,.
 }

As explained in section~\ref{DIAGS3}, the $\mathbb{S}^{n \ell m}$ are gradients of the usual scalar harmonics:
 \es{SFormula}{
  {\mathbb{S}}^{n \ell m}(\chi, \theta, \phi) = \frac{d Y_{n \ell m}(\chi, \theta, \phi)}{\sqrt{n^2 -1}}  \,,
 }
and they are the only closed forms in the decomposition \eqref{omegaDecomp}.   The co-closed forms $\mathbb{V}^{n \ell m}_{L}$ and $\mathbb{V}^{n \ell m}_{R}$ can be recast into the symmetric and antisymmetric combinations
 \es{GotVW}{
  \mathbb{V}^{n \ell m} &= \frac{\mathbb{V}^{n \ell m}_{L} + \mathbb{V}^{n \ell m}_{R}}{\sqrt{2}}  \,, \qquad
   \mathbb{W}^{n \ell m} = \frac{\mathbb{V}^{n \ell m}_{L} - \mathbb{V}^{n \ell m}_{R}}{\sqrt{2}}  \,,
 }
which by virtue of \eqref{Normalizations} are also orthonormal.  Then
 \es{NewVWFormulas}{
  \mathbb{V}^{n \ell m}(\chi, \theta, \phi) &=  *d \left( \frac{\sin^{\ell+1} \chi \, \Phi_{n \ell} (\chi)}{\sqrt{n^2 \ell(\ell+1) a_{n \ell}}}   *_2 dY_{\ell m}(\theta, \phi) \right) \,, \\
    \mathbb{W}^{n \ell m}(\chi, \theta, \phi) &=
   \frac{\sin^{\ell + 1} \chi\, \Phi_{n \ell}(\chi)}{\sqrt{\ell (\ell + 1) a_{n \ell}}}
   * (d\chi \wedge  dY_{\ell m}(\theta, \phi)) \,,
 }
where $*_2$ denotes the Hodge dual on $S^2$ with the standard line element.  These expressions exhibit $\mathbb{V}^{n \ell m}$ and $\mathbb{W}^{n \ell m}$ explicitly as co-closed one-forms.

When we evaluate the sums in equations like \eqref{seigenvalues} we need to know what happens to the one-forms close to the North pole at $\chi = 0$.  A simple analysis of eqs.~\eqref{SFormula} and \eqref{NewVWFormulas} gives:
 \es{Asymp}{
  \abs{\mathbb{S}^{n \ell m}(\chi, \theta, \phi)}^2 &= O(\chi^{2 \ell - 2}) \,,
   \qquad \abs{\mathbb{V}^{n \ell m}(\chi, \theta, \phi)}^2 = O(\chi^{2 \ell - 2}) \,, \\
   \qquad \abs{\mathbb{W}^{n \ell m}(\chi, \theta, \phi)}^2 &= O(\chi^{2 \ell})
 }
as $\chi \to 0$.  The only harmonics that are non-zero at $\chi = 0$ are therefore $\mathbb{S}^{n \ell m}$ and $\mathbb{V}^{n \ell m}$ with $\ell = 1$ and $m = -1, 0, 1$.

\bibliographystyle{ssg}
\bibliography{gauge}

\end{document}